\documentclass[12pt]{article}

\pdfoutput=1

\usepackage{url}
\usepackage{amssymb}
\usepackage{graphicx}
\usepackage[normalem]{ulem}
\usepackage[utf8]{inputenc} 
\usepackage{color}

\def\lbldef#1#2{\expandafter\gdef\csname #1\endcsname {#2}}

\def\href#1#2{#2}

\begin{document}

\begin{titlepage}

\title{\vspace*{-2.0cm}
\hfill {\small FTUAM-15-43}\\
\vglue -0.3cm
\hfill {\small IFT-UAM/CSIC-15-125} \vskip 0.2cm
\bf\Large
Neutrino mass limits: robust information from \\the power spectrum of galaxy surveys}

\author{
Antonio J. Cuesta$^a${\let\thefootnote\relax\footnote{author list alphabetized}}\thanks{email: \tt ajcuesta@icc.ub.edu},~~Viviana Niro$^b$\thanks{email: \tt viviana.niro@uam.es},~~Licia Verde$^{acde}$\thanks{email: \tt liciaverde@icc.ub.edu}\\ \\
$^a${\normalsize \it Institut de Ci{\`e}ncies del Cosmos (ICCUB), Universitat de Barcelona (IEEC-UB),} \\
{\normalsize \it Mart{\'\i} i Franqu{\`e}s 1, E08028 Barcelona, Spain}\\
$^b${\normalsize \it Departamento de F\'isica Te\'orica, Universidad Aut\'onoma de Madrid, and Instituto}\\
{\normalsize \it de F\'isica Te\'orica UAM/CSIC, Calle Nicol\'as Cabrera 13-15, Cantoblanco, E-28049 Madrid, Spain} \\
$^c${\normalsize \it  ICREA (Instituci\'o catalana de recerca i estudis avan\c{c}ats)} \\
$^d${\normalsize \it Radcliffe Institute for Advanced Study \& ITC, } \\
{\normalsize \it Harvard--Smithsonian Center for Astrophysics, Harvard University, MA 02138, USA} \\
$^e${\normalsize \it  Institute of Theoretical Astrophysics, University of Oslo, 0315 Oslo, Norway}
} 
\date{\today}
\maketitle
\thispagestyle{empty}

\vspace*{-1.0cm}
\begin{abstract}
\noindent
We present cosmological upper limits on the sum of active neutrino masses using large-scale power spectrum
data from the WiggleZ Dark Energy Survey and from the Sloan Digital Sky Survey - Data Release 7 (SDSS-DR7) sample of Luminous Red Galaxies (LRG).
Combining measurements on the Cosmic Microwave Background temperature and polarisation anisotropies  by the Planck satellite
together with WiggleZ power spectrum results in a neutrino mass bound of 0.37~eV  at 95\%~C.L.,
while replacing WiggleZ by the SDSS-DR7 LRG power spectrum, the 95\%~C.L. bound on the sum of neutrino masses is 0.38~eV.
Adding Baryon Acoustic Oscillation (BAO) distance scale measurements, the neutrino mass upper limits greatly improve, since BAO data  break degeneracies
in parameter space. Within a $\Lambda$CDM model, we find an upper limit of 0.13~eV (0.14~eV)  at 95\%~C.L., when using  SDSS-DR7 LRG (WiggleZ) together with BAO and
Planck. The addition of BAO data makes the neutrino mass upper limit robust, showing only a weak dependence
on the  power spectrum  used. 
We also quantify the dependence of neutrino mass limit reported here on the CMB lensing information.
The tighter upper limit (0.13~eV) obtained with  SDSS-DR7 LRG is very close to that  recently obtained using Lyman-alpha clustering data, yet uses a completely different
 probe and redshift range,  further supporting the robustness of the  constraint.
 This constraint puts under some pressure the inverted mass hierarchy and favours the normal hierarchy.
\end{abstract}

\end{titlepage}

\section{Introduction}

Neutrino oscillation experiments, including solar, atmospheric, and reactor experiments, require that neutrinos have mass 
(Nobel Prize in Physics 2015; see for example Ref.~\cite{Gonzalez-Garcia:2014bfa,Bergstrom:2015rba}). 
However, the neutrino mass hierarchy, normal or inverted, as well as the sum of the three active neutrino masses, 
are quantities that are still unknown. 

Cosmology provides important information on the sum of neutrino masses, $M_{\nu}$, since this quantity 
affects the expansion rate of the Universe and the way large-scale structures form and evolve. 
 
Thus, an interplay between laboratory experiments and cosmology on 
neutrino masses can provide significant information on neutrino 
physics, e.g., Ref.~\cite{JKPGV, GonzalezGarcia:2010un,Gerbino:2015ixa} and refs. therein.  
It is from this interplay for example that cosmology  helps determining  also the mass hierarchy. Should the cosmological constraints of the sum of the masses bound $M_{\nu}<0.1$~eV at reliable significance level, this would indicate a normal mass hierarchy and exclude the inverted one. This is why reaching the 0.1 eV precision (and accuracy) in $M_{\nu}$ is a highly desirable target and would have profound implications beyond cosmology. Considering the analysis of Ref.~\cite{Gonzalez-Garcia:2014bfa}, 
the current minimum value allowed for an inverted 
hierarchy is $M_{\nu}=0.0982 \pm  0.0010$ eV (68\% C.L.). 
If the lightest state, $m_3$, is massless, then $m_2^2$ = $|\Delta m^2_{32}|$ and $m_1^2 = |\Delta m^2_{32}| - \Delta m^2_{21}$ with 
the best-fit for $|\Delta m^2_{32}|$ and $\Delta m^2_{21}$ equal respectively to $2.449 \times 10^{-3}~{\rm eV}^2$ and  $7.5 \times 10^{-5}~{\rm eV}^2$. The 
sum of the masses, $M_\nu$, is thus $0.0982$~eV. Following this 
way of reasoning, it is possible to find the error associated to $M_\nu$ at 68\%~C.L..

Neutrino masses affect on the Cosmic Microwave Background (CMB) anisotropies through the Early Integrated Sachs Wolfe Effect, 
and they influence gravitational lensing measurements~\cite{Lesgourgues:2005yv}. Massive neutrinos, moreover, lead to a suppression on the matter power spectrum at  small scales.  For this reason, measurements of the full shape of the matter power spectrum are of great 
importance for neutrino physics since they are able to put tight constraints on 
the sum of neutrino masses. 

This has been shown previously in the literature starting with the pioneering work of \cite{Hu:1997mj}, see \cite{Lesgourgues:2006nd} for a review. 
Recently, using the WiggleZ (WZ) Dark Energy Survey  galaxy power spectrum~\cite{Riemer-Sorensen:2013jsa}, it has been shown that WZ combined with Planck~2013 data 
can provide tight constraints on the sum of neutrino masses.  Before WZ, the state-of-the art galaxy survey for neutrino mass analysis was the SDSS DR7, used e.g. in \cite{Reid:2009nq} or its photometric counterpart in~\cite{Thomas:2009ae}.
To date the  strongest constraint on $M_{\nu}$ is provided by the joint analysis of CMB, BAO and Lyman-$\alpha$ forest data \cite{Palanque-Delabrouille:2014jca, Palanque-Delabrouille:2015pga}, $M_{\nu}<0.12$~eV (95\% C.L.), 
reaching tantalisingly close to the 0.1 eV target. The next generation of cosmological surveys have the statistical power to detect the signature of non-zero $M_{\nu}$  even if it had the  minimum value allowed by oscillation measurements and the mass  hierarchy was normal e.g.,\cite{Carbone:2010ik, Audren:2012vy,Hamann:2012fe,Font-Ribera:2013rwa}.

All these analyses use the power spectrum of dark matter tracers  and thus rely on the assumption that it can be used as a  proxy for the dark matter power spectrum itself. Galaxies are not expected to Poisson-sample the matter distribution; the galaxy--dark matter connection is  complicated and not yet well understood; galaxies (but also  the intergalactic medium) are known  to be biased tracers of the dark matter density field, and  different tracers  are known (see pioneering work by~\cite{1976ApJ...208...13D}) to cluster differently.
Galaxy bias  probably represent  the  main limitation in extracting constraints on neutrino masses from galaxy surveys, which can introduce potentially crippling systematic errors on this parameter as the upper limits shrink towards the 0.1 eV target.
In Ref.~\cite{2010MNRAS.409.1100S} the authors address this issue by considering neutrino mass limits from the clustering of red and blue galaxies of the SDSS DR7 main sample. They find different upper limits between the two  samples (by $0.22$ eV) when both are combined with WMAP 5 years  CMB power spectrum data, which they interpret as broad agreement for their data set but a hint of a possible issue with galaxy bias for future data. 

In Ref.~\cite{Zhao:2012xw}, instead, the authors considered the limits on neutrino mass obtained from WMAP 7 years CMB power spectrum data, supernova data, BAO data, and the galaxy power spectrum of the SDSS-III BOSS Data Release 9 (DR9) CMASS galaxy sample, obtaining a bound of the order of 0.340~eV at 95\%~C.L. Our analysis (and thus our upper bounds) differs on several points: we use the Planck temperature and polarisation data, a new set of BAO data, and the {\it halo} $P(k)$ from DR7, which reduces uncertainties due to non-linear effects. 

Here we perform a more extreme test by considering luminous red galaxies (LRG in DR7) and  emission line galaxies (WZ). LRG are mostly passively evolving, massive galaxies thought to inhabit preferentially the centre of cluster-size dark matter halos~\cite{Eisenstein2001, Reid09}. The WZ galaxies are star-forming blue galaxies which tend to avoid the densest regions (hence  avoid the centre of cluster-size dark matter halos)~\cite{2010MNRAS.401.1429D}. Moreover the power spectrum extracted from the LRG catalogue is an estimate of the power spectrum of the massive halos hosting the  LRG galaxy population. Thus not only the bias of the two tracers are very different but also the sensitivity of their power spectra to non-linearities and non-linear redshift space distortions. 
Because of the extremely different properties of the tracers, agreement in the constraints obtained would signify that any systematic effects due to galaxy bias (and non-linearities) are below the  reported error-bars. 

In this paper  not only for the first time we compare  the constraints on neutrino masses obtained from the shape of the power spectrum of SDSS DR7 LRG galaxies and that of WZ galaxies, but also consider state-of-the art data from  CMB  observations of the Planck satellite (Planck~2015)~\cite{Ade:2015xua} and BAO data. 
Note that precision of the Planck~2015 data has also greatly improved the sensitivity to 
test perturbations in the cosmic neutrino background, see for example 
the analysis in Refs.~\cite{Audren:2014lsa,Sellentin:2014gaa,Ade:2015xua} (for the case of self-interacting neutrinos 
see instead the discussion in Refs.~\cite{Cyr-Racine:2013jua,Oldengott:2014qra}).

This paper is organised as follows: In Sect.~\ref{sec:num} we present our methodology, in 
particular we describe the parameterisation adopted and the data sets considered. 
In Sect.~\ref{sec:res} we present our results on the upper limit on the sum of neutrino masses from different  data set combinations. We discuss our findings and  compare them with other  constraints  presented in the literature. Finally we conclude in  Sect.~\ref{sec:conc}.

\section{\label{sec:num}Methods and data}

\subsection{\label{sec:cosmo}Model parameterisation and fitting procedure}

 We consider the six parameters describing the $\Lambda$CDM model 
with the addition of the sum of active neutrino masses. We perform standard Bayesian inference 
using  the following parameters:  
\[
\{\omega_b, \omega_{\rm{cdm}}, 100\Theta_s, \ln(10^{10}A_s), n_s, \tau_{\rm{reio}}, \sum m_\nu \} \,.
\]
The first six cosmological parameters denote the physical baryon density ($\omega_b \equiv \Omega_b h^2$), the 
physical cold dark matter density ($\omega_{\rm{cdm}} \equiv \Omega_{\rm{cdm}} h^2$), the ratio between the sound 
horizon and the angular diameter distance at decoupling ($\Theta_s$), the
amplitude ($A_s$) of the power spectrum of primordial fluctuations at the pivot scale $k_* = 0.05/$Mpc, the power spectrum index ($n_s$) of the primordial density fluctuations, 
the optical depth to reionisation ($\tau_{\rm{reio}}$), and the sum of the three active neutrino masses ($\sum m_\nu \equiv M_\nu$). 

We assume an uniform prior for all these parameters. 
We do not enforce any lower or upper limit for all the parameters, except for $m_\nu$, assumed to be positive,  
and for $\tau_{\rm{reio}}$. For the latter  we enforce a lower bound of 0.04, since lower values of $\tau_{\rm{reio}}$ would be inconsistent 
with observations on the Gunn--Peterson effect, see e.g., Ref.~\cite{Caruana:2013qua}, but this prior does not affect our conclusions or our bounds. 
Unless otherwise stated we will use the standard, official and recommended likelihood for each of the data sets. 
The officially released likelihood software depends also on extra (nuisance) parameters which should be (and are) 
marginalised over. For example the Planck (CMB data) likelihood depends on several parameters describing the residual 
foreground contamination, and the galaxy surveys likelihood depends on parameters describing residual 
effects of e.g., bias and non-linearities. More detailed information can be found in the references relative to the 
specific data sets (see sec.~\ref{sec:data}) and in the accompanying notes to the officially released software (see below). 

In principle, since the neutrino mass upper limit can constrain the specific neutrino hierarchy (i.e., normal or 
inverted), there could be a weak dependence of the results on the adopted hierarchy, see  Ref.~\cite{Riemer-Sorensen:2013jsa} and  also the discussion in Ref.~\cite{Palanque-Delabrouille:2015pga}, 
in which specific analyses considering different neutrino ordering have been carried out. 
In our analysis, we consider three degenerate massive neutrinos, setting a lower bound equal to zero on $M_\nu$. This choice is
 analogous to the one considered in the analysis of the Planck collaboration~\cite{Ade:2015xua}. 
For the expected error-bars, this approximation introduces negligible errors. 
In this way, we can derive an independent upper limit set by cosmology on the total mass of active 
neutrinos, without considering the information reported by neutrino oscillation experiments. However, for  future surveys  with reduced error-bars the adopted hierarchy matters, see e.g., \cite{Carbone:2010ik, Kitching08} and refs. therein. 

We use the Cosmic Linear Anisotropy Solver Software (CLASS)~\footnote{\url{http://class-code.net}} presented in 
Ref.~\cite{Lesgourgues:2011re} for the solution of Boltzmann equations 
and the calculation of Cosmic Microwave Background and large scale structure observables. 
For the parameter inference we use the public code {\sc Monte
Python}~\footnote{\url{http://baudren.github.io/montepython.html}} described in Ref.~\cite{Audren:2012wb}. This is based on a Metropolis Hastings algorithm and it uses a 
Cholesky decomposition in order to handle a large number of nuisance parameters~\cite{Lewis:2013hha}. 

Note that we will not consider extended cosmological models and their effect on the upper bound for the 
three active neutrino masses. In these more general cases, for example, relaxing spatial flatness, 
varying $N_{\rm eff}$ or adding a massive sterile neutrino $m_{\nu,{\rm sterile}}^{\rm eff}$ --see 
e.g., Refs.~\cite{Archidiacono:2013xxa,Bergstrom:2014fqa,Gariazzo:2013gua,Mirizzi:2013gnd, V2013, F2013} for  constraints on sterile neutrinos from cosmology--, the bounds on the sum of 
the three active neutrino masses are, in general, less constraining. To understand how these different assumptions might 
affect the neutrino mass bounds, we 
refer to Ref.~\cite{Ade:2015xua}, see also Ref.~\cite{GonzalezGarcia:2010un} for a previous study on this topic. 
Finally, constraints on neutrino masses could change also in the presence of cosmological axions. 
For a recent analysis on this topic, using the Planck~2015 data release, see Ref.~\cite{DiValentino:2015wba}, see also 
Ref.~\cite{Archidiacono:2013cha} for a previous study using Planck~2013 data and see Ref.~\cite{Archidiacono:2015mda} for 
the potential of future surveys such as Euclid. 
The impact of dark energy on neutrino upper limits has been recently presented in Ref.~\cite{Zhang:2015uhk}.

\begin{figure}[t!]
\centering
\includegraphics[width=0.45\textwidth]{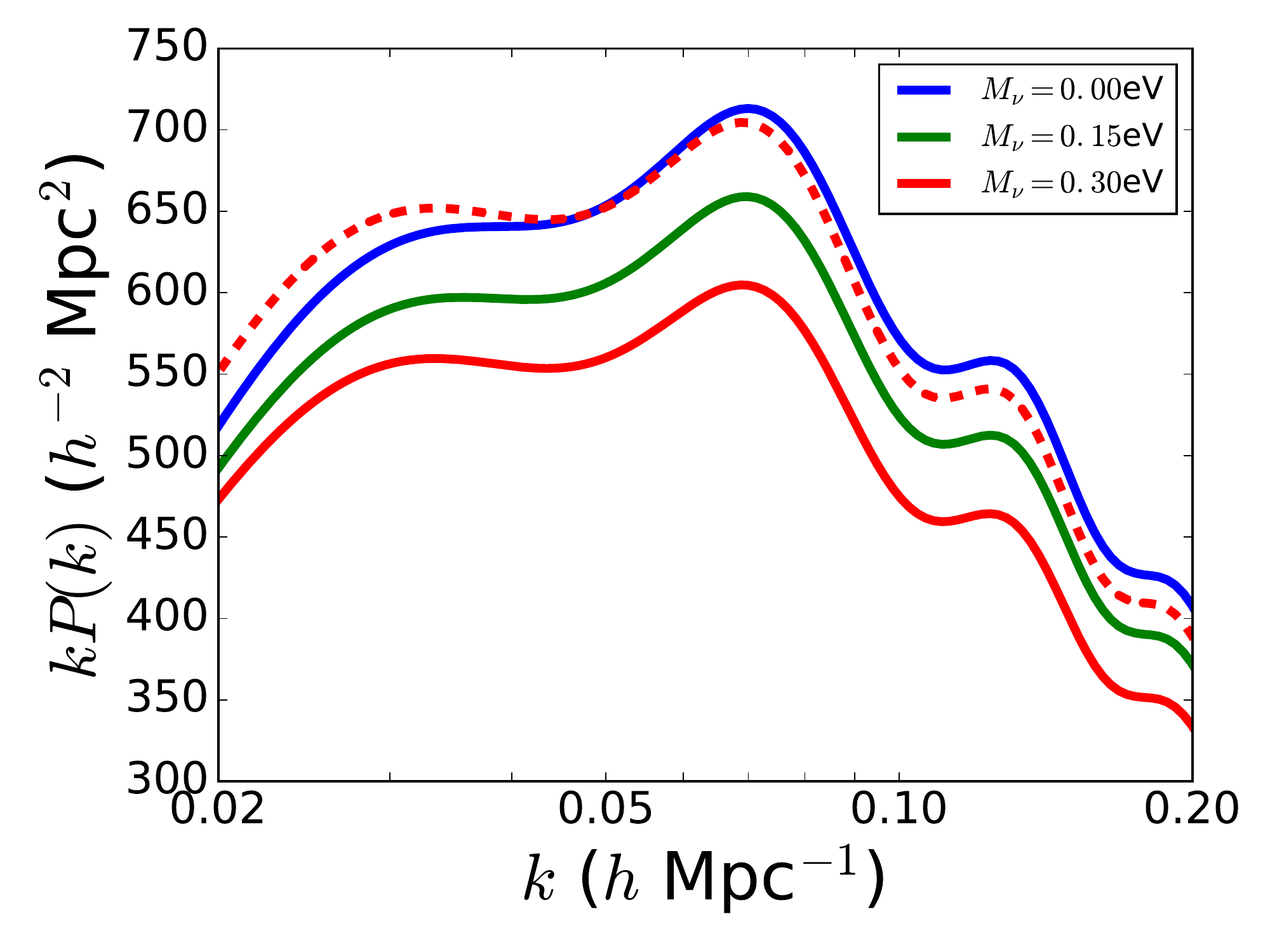}
\includegraphics[width=0.45\textwidth]{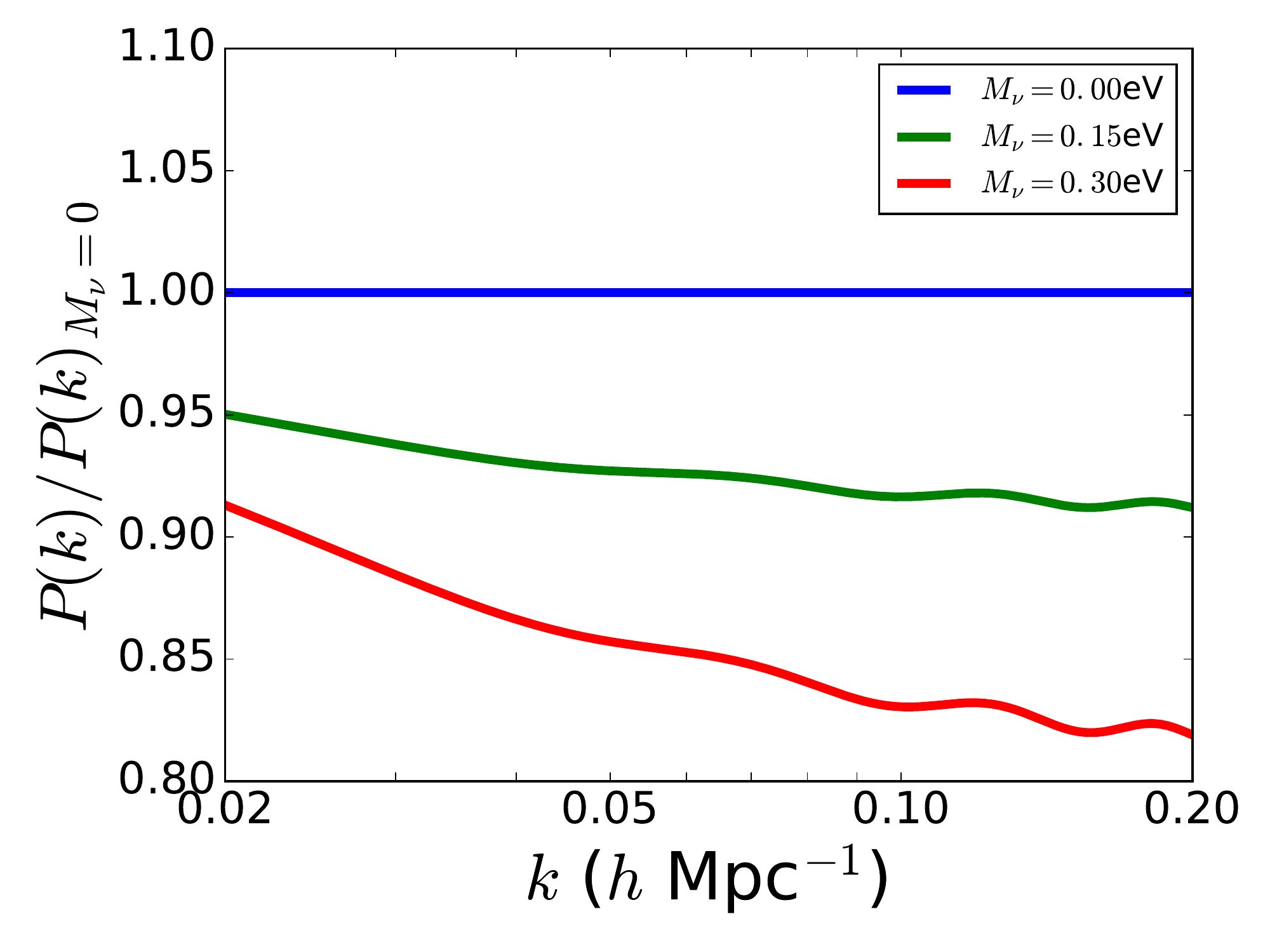}
\caption{Effect (absolute on the left panel and relative to the massless neutrino case on the right panel) of neutrino mass on the non-linear matter power spectrum at $z=0$ for the $k$-range used in this analysis,  prior to applying the window function of the survey. Shown are the power spectrum for massless neutrinos (blue),  for the neutrino mass  close to the CMB+BAO+$P(k)$ bound in this paper (green), and for the neutrino mass close to the bound given by current CMB+$P(k)$ data (red) while $\Omega_m$, and all other cosmological parameters (except for $\omega_{\rm{cdm}}$) are kept fixed. The dashed red line is the red line renormalised (akin to a scale-independent bias factor) to match the blue one at $k=0.05$ $h$ Mpc$^{-1}$. }
\label{fig:pk_mnu}
\end{figure}

\subsection{\label{sec:data}Data sets  and data modelling}
In the following we outline the data sets used. 
\vspace{0.1cm} \\

\noindent \textbf{Cosmic Microwave Background:} \\
We use information on the CMB from the Planck~2015 data releases. 
When analysing the recent Planck~2015 measurements of temperature and polarisation 
anisotropies~\cite{Ade:2015xua}, we consider a combination of
the likelihood at $l\geq30$ using TT, TE and EE power spectra and the Planck low-multipole temperature+polarisation likelihood at  $l\le 29$. This combination has been referred as 
Planck TT,TE,EE+lowP in Ref.~\cite{Ade:2015xua}. In the following, we will use the nomenclature of ``CMB15'' 
to refer to these combination of data. We also include the Planck~2015 lensing likelihood, as constructed from 
measurements of the power spectrum of the lensing potential $C_l^{\phi \phi}$~\cite{Ade:2015zua}. 
\vspace{0.2cm} \\

\noindent \textbf{Power spectrum $P(k)$:}\\
We use the power spectrum of the halo density field derived by \cite{Reid:2009xm} from a sample of Luminous Red Galaxies in the Sloan Digital Sky Survey DR7 (LRG). 
The scales probed by this measurement range from $k=$0.02 $h~$Mpc$^{-1}$ to 0.2 $h~$Mpc$^{-1}$. 

We have rewritten the original likelihood code~\footnote{\url{http://lambda.gsfc.nasa.gov/toolbox/lrgdr/}} into Python so that we could interface it through \textsc{MontePython}. The treatment of non-linearities is analogous to the original code, in which the linear matter power spectrum is fed into \textsc{Halofit} to compute non-linear effects, and a smooth polynomial derived from $N$-body simulations is used to account for additional scale-dependent effects between the halo power spectrum and matter power spectrum. However, since in our case we are using an updated, more accurate version of Halofit \cite{Takahashi:2012em, BirdNus} which models the effects on massive neutrinos in the non-linear corrections, we recomputed the non-linear $P(k)$ ratios for their fiducial cosmology and replaced the original files with these updated templates. As in the original code, we also include the (redshift-dependent) effect of BAO smoothing due to non-linear evolution. In order to do so, we generate the no-wiggle power spectrum \cite{EH} using the code \textsc{CosmoloPy} \footnote{\url{http://roban.github.com/CosmoloPy/}} modified to implement a more accurate approximation for massive neutrinos as presented in \cite{2008PhRvD..77f3005K}.

We use also power spectrum data from the WiggleZ Dark Energy Survey~\cite{Parkinson:2012vd}, that 
consists of galaxies in a redshift range $z<1.0$. For detailed information on the WZ survey and data 
releases, see also Refs.~\cite{Blake:2010xz, Drinkwater:2009sd} 
We restrict the analysis to  $k<k_{\rm{max}}=0.2 \ h$~Mpc$^{-1}$, to 
minimise the uncertainties in the modelling of the non-linear regime as shown in  Ref.~\cite{RiemerSorensen:2011fe, Riemer-Sorensen:2013jsa}. 
The WiggleZ modelling is similar to the LRG modelling except for the BAO smoothing, which is not implemented for WiggleZ.

\begin{figure}[t!]
\centering
\includegraphics[width=0.7\textwidth]{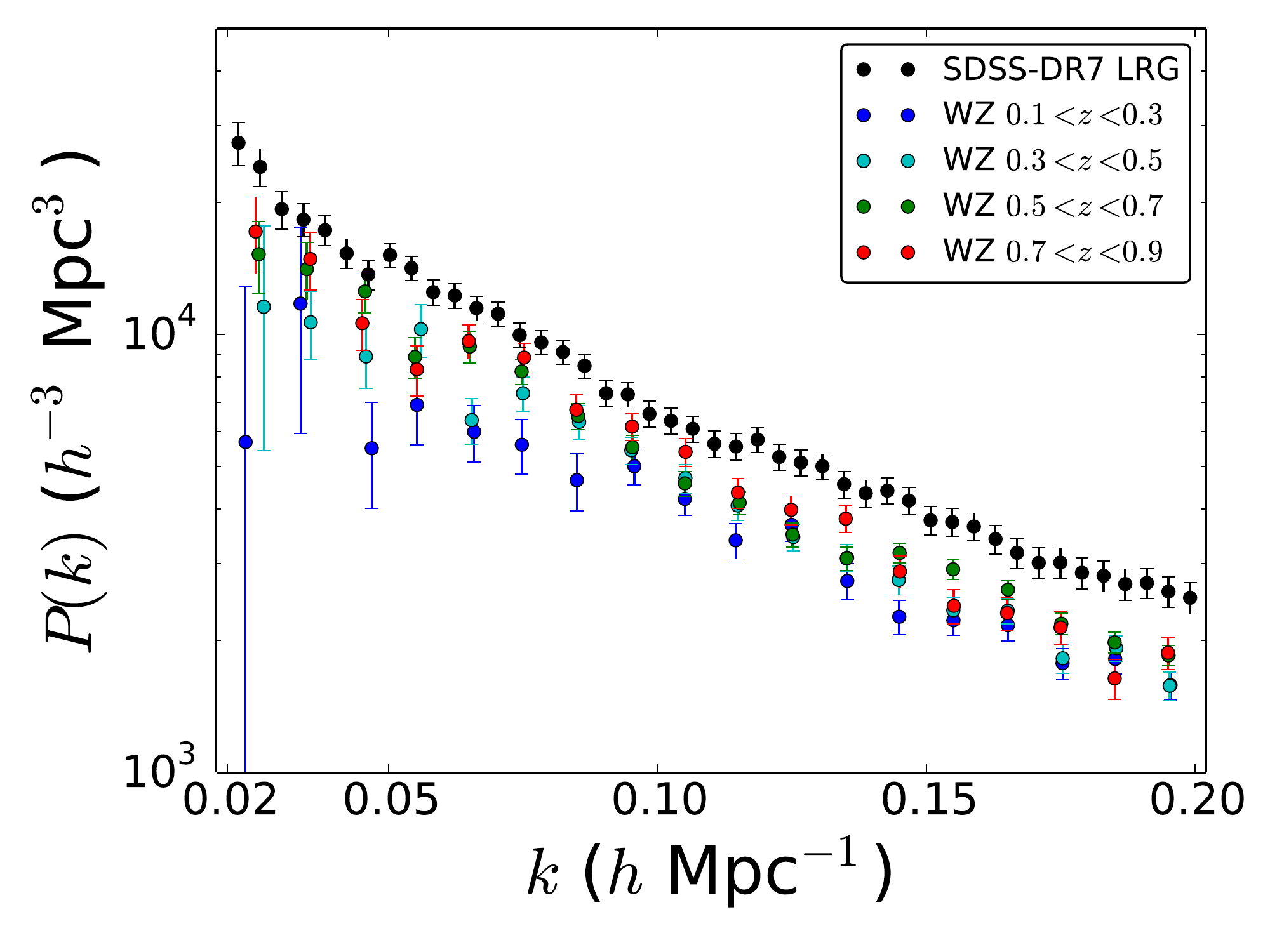}
\caption{LRG and WZ data sets used. The LRG data are at $z\sim 0.35$ (but normalised to $z=0$) and WZ is separated in  four redshift bins  at effective redshifts $z=0.22, 0.41, 0.6, 0.78$, only band powers with $k<0.2$ $h$ Mpc$^{-1}$ are used here. A visual comparison of the two data sets might be misleading since not only the galaxy samples but also   the window functions of the two surveys are different. }
\label{fig:wz_lrg_pk}
\end{figure}

The non-linear matter power spectrum at $z=0$ over the range of scales used here is shown in Fig.~\ref{fig:pk_mnu} for several values of $M_{\nu}$.
The power spectra for  LRG and WZ  data  are shown in Fig.~\ref{fig:wz_lrg_pk}.

The treatment of real-world effects such as non-linearities or redshift-space distortions is the standard one 
as coded in the officially-released likelihood codes. In particular, for redshift space distortions, since we consider the angle-averaged power 
spectrum, the effect of redshift space distortions is its mean, angle-averaged effect. In the linear regime this is equivalent to a scale-invariant 
boost of power which gets automatically marginalised over when marginalising over the (unknown) bias amplitude. (Note that treating the overall amplitude as a nuisance parameter marginalises automatically over other physical effects which affect the overall amplitude such as super-survey effects).   
Non-linear redshift space distortions do induce a scale-dependent effect, -- sometimes referred to as Fingers-of-God -- which 
is relatively small in the scales of interest. As explained in Reid et al. 2010~\cite{Reid:2009xm} and references therein, for LRG, 
since a halo field was reconstructed from the galaxy distribution, the Fingers-of-God arise from the virial motions of tracers within 
the same halo. By using, instead of individual galaxies,  the estimated position of the halo centre of mass, this effect disappears. 
Small residual scale-dependent corrections to the matter power spectrum shape are marginalised over via nuisance parameters. 
For WZ the higher redshift of the sample, reduces the impact of non-linearities at such large scales. Nevertheless the WZ team carefully calibrated 
the marginalisation of such scale-dependent effects in Refs.~\cite{RiemerSorensen:2011fe,Riemer-Sorensen:2013jsa}, where they explore several 
different modelling approaches, quantified their performance on mock survey catalogues and justified their selection of the optimal method.

Refs.~\cite{Brandbyge:2010ge,Ichiki:2011ue,Castorina:2013wga,Castorina:2015bma}  highlighted a modification of the halo mass function and the halo-bias 
due to the presence of massive neutrinos. 
The first effect, that is present especially for high neutrino masses, is not relevant for our analysis, since 
for WZ there are no halos involved, while for LRG a small modifications of the shape of the halo mass function are 
fully sub-dominant to other effects, which are marginalised over. 
The second effect is the (small) scale dependence of the halo bias induced by 
neutrino masses when compared to standard (massless neutrinos) $\Lambda$CDM. 
The scale-dependent effect is present especially for neutrino masses 
of about 0.5 eV and above, and for high redshifts. 
These high values of masses are constrained by CMB data 
and the scale dependence of the halo bias effect is smaller for the values of masses which are of interest in 
this current analysis, see for example Fig.~12 in Ref.~\cite{Castorina:2015bma}. 
Moreover, this effect does decrease at lower redshifts and note that the effective redshift of the present sample 
is lower than the lowest redshift presented in the above reference. 
The current approach in the LRG analysis already marginalises generously over a possible 
halo scale-dependent bias, and thus safely over this effect. 
\vspace{0.2cm} \\

\noindent \textbf{Baryon Acoustic Oscillations:} \\
BAO measurements  greatly help to lift  cosmological parameters degeneracies  (in particular by ruling out small values of $H_0$ allowed by the CMB-only case) which affect constraints obtained from CMB data. 
Specifically, we use the recent determinations of $D_V/r_{\rm drag}$ as reported by 
the analysis of the Six-degree-Field Galaxy Survey (6dFGS)~\cite{Beutler:2011hx}, the SDSS Main Galaxy Sample (MGS)~\cite{Ross:2014qpa}, 
the Baryon Oscillation Spectroscopic Survey (BOSS) LOWZ and CMASS samples~\cite{Anderson:2013zyy}, where we use the isotropic BAO measurement in the case of CMASS. \footnote{In this case, the combination of CMB15 with these BAO data sets results in a neutrino mass bound of 0.14~eV, slightly more constraining than the one reported by \cite{Ade:2015xua} for the CMB15+BAO combination (0.17~eV). We found that this improvement is mostly attributed to using the \textit{isotropic} (rather than the anisotropic) BAO measurement of CMASS.}
Note that the BAO data used in our analysis are  different from the one used by the WiggleZ team in Ref.~\cite{Riemer-Sorensen:2013jsa} 
and from those used by the Planck collaboration in the 2013 release~\cite{PlanckXVI}. 
When we combine BAO information with the full matter power spectrum data from the SDSS-DR7 LRG, 
we have not added the SDSS-MGS BAO data set of Ref.~\cite{Ross:2014qpa}, to avoid possible double counting of information. We did not use in this study the information from Lyman-$\alpha$ forest BAO data~\cite{Font-Ribera:2013wce}. A summary of the values used in our analysis for the BAO measurements can be found in 
Table~\ref{tab:bao}. 

\begin{table}
\centering
\begin{tabular}{ c || c  } 
Survey & Measurement  \\ \hline \hline
6dFGS & $r_{\rm drag}/D_V (z=0.106)$: \, $0.327\pm0.015$~\cite{Beutler:2011hx}\\ \hline
SDSS-MGS & $D_V (z=0.15)/r_{\rm drag}$: \, $4.47\pm0.16$~\cite{Ross:2014qpa} \\ \hline
BOSS-LOWZ & $D_V (z=0.32)/r_{\rm drag}$: \, $8.47\pm0.17$~\cite{Anderson:2013zyy}\\ \hline
BOSS-CMASS & $D_V (z=0.57)/r_{\rm drag}$: \, $13.77\pm0.13$~\cite{Anderson:2013zyy}\\ \hline
\end{tabular}
\caption{BAO cosmic distance scale measurements used in this paper.}
\label{tab:bao}
\end{table}

\section{\label{sec:res}Results}

\begin{figure}[t!]
\centering
\includegraphics[width=0.45\textwidth]{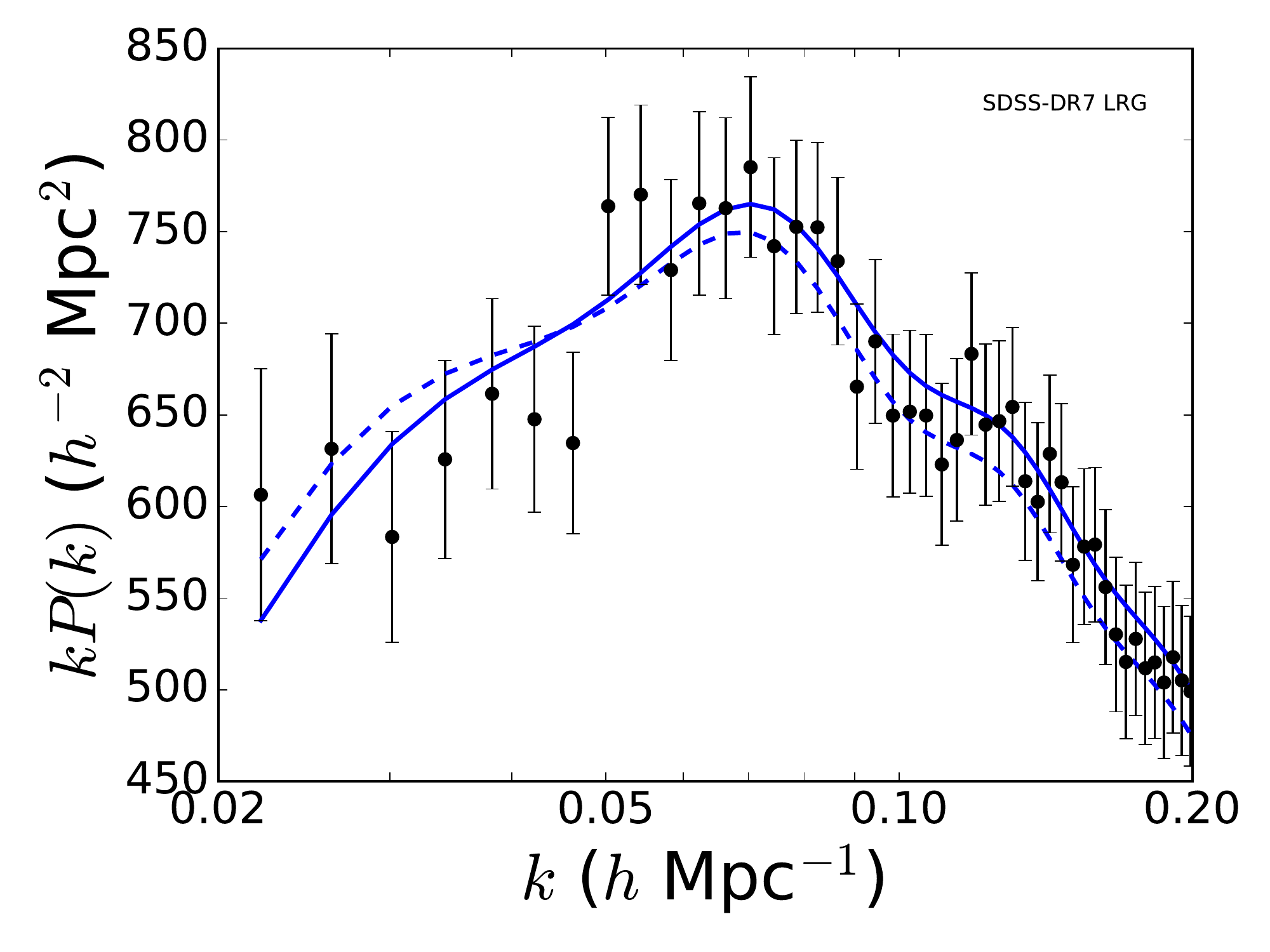}
\includegraphics[width=0.45\textwidth]{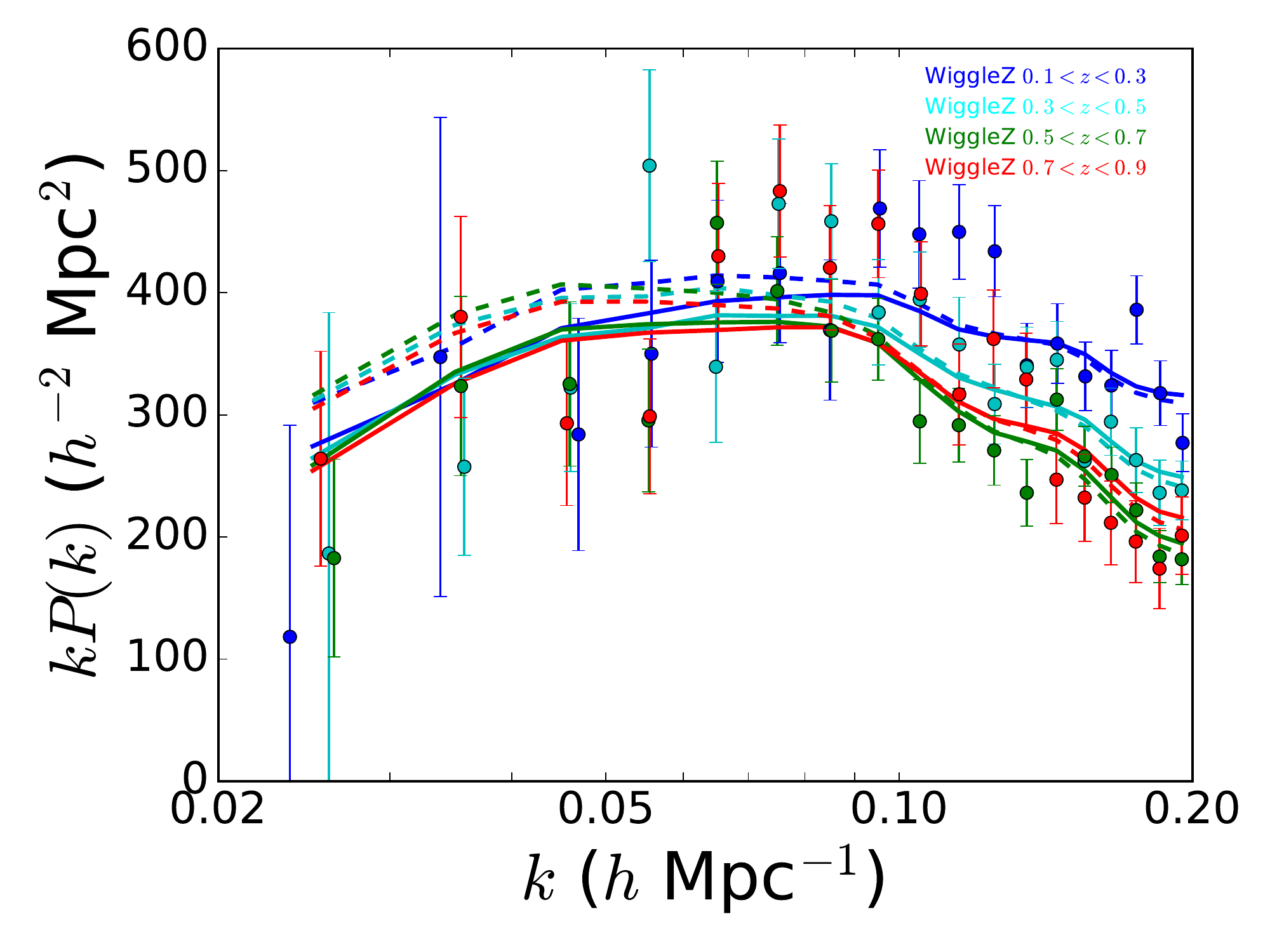}
\includegraphics[width=0.45\textwidth]{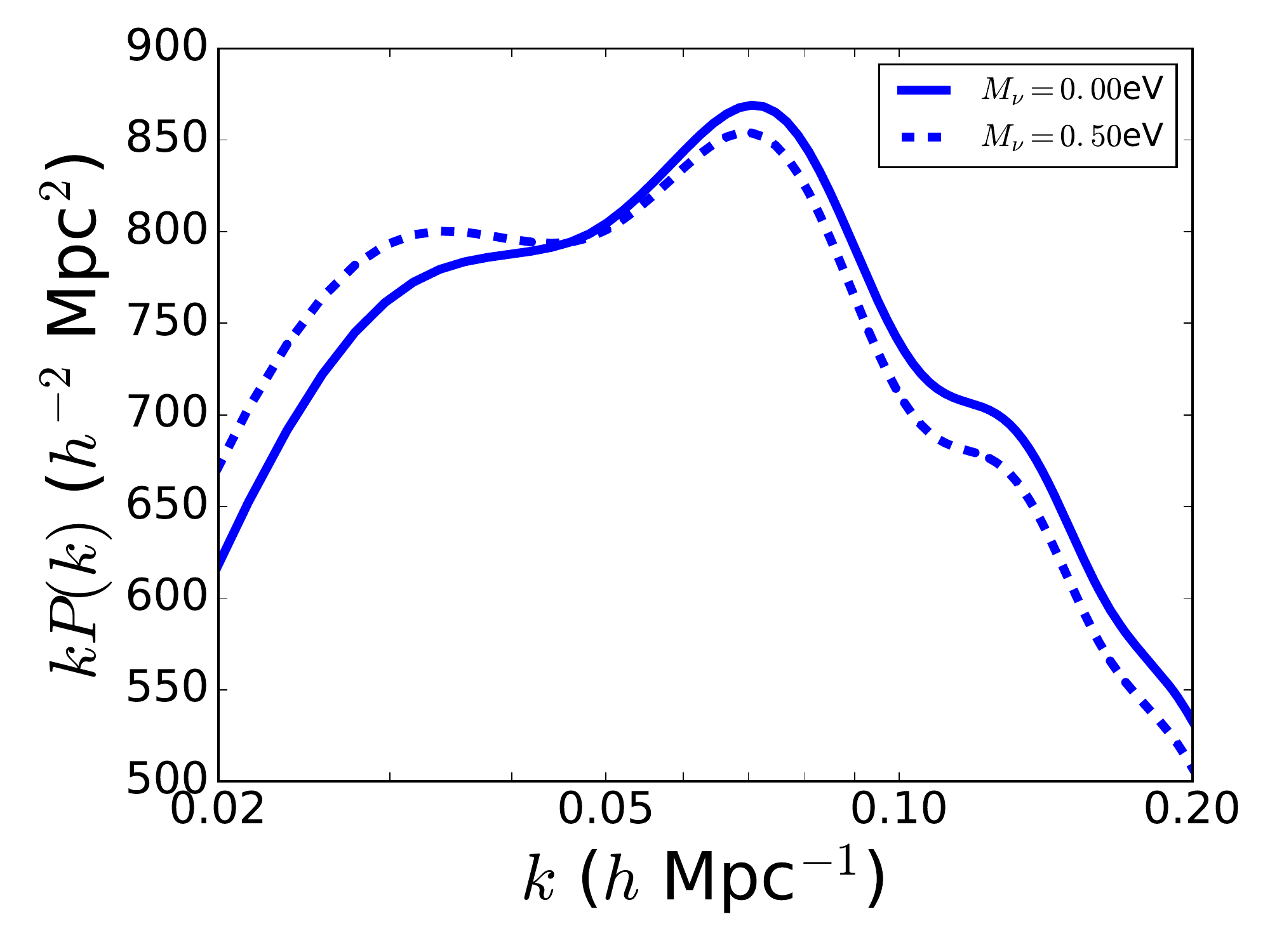}
\includegraphics[width=0.45\textwidth]{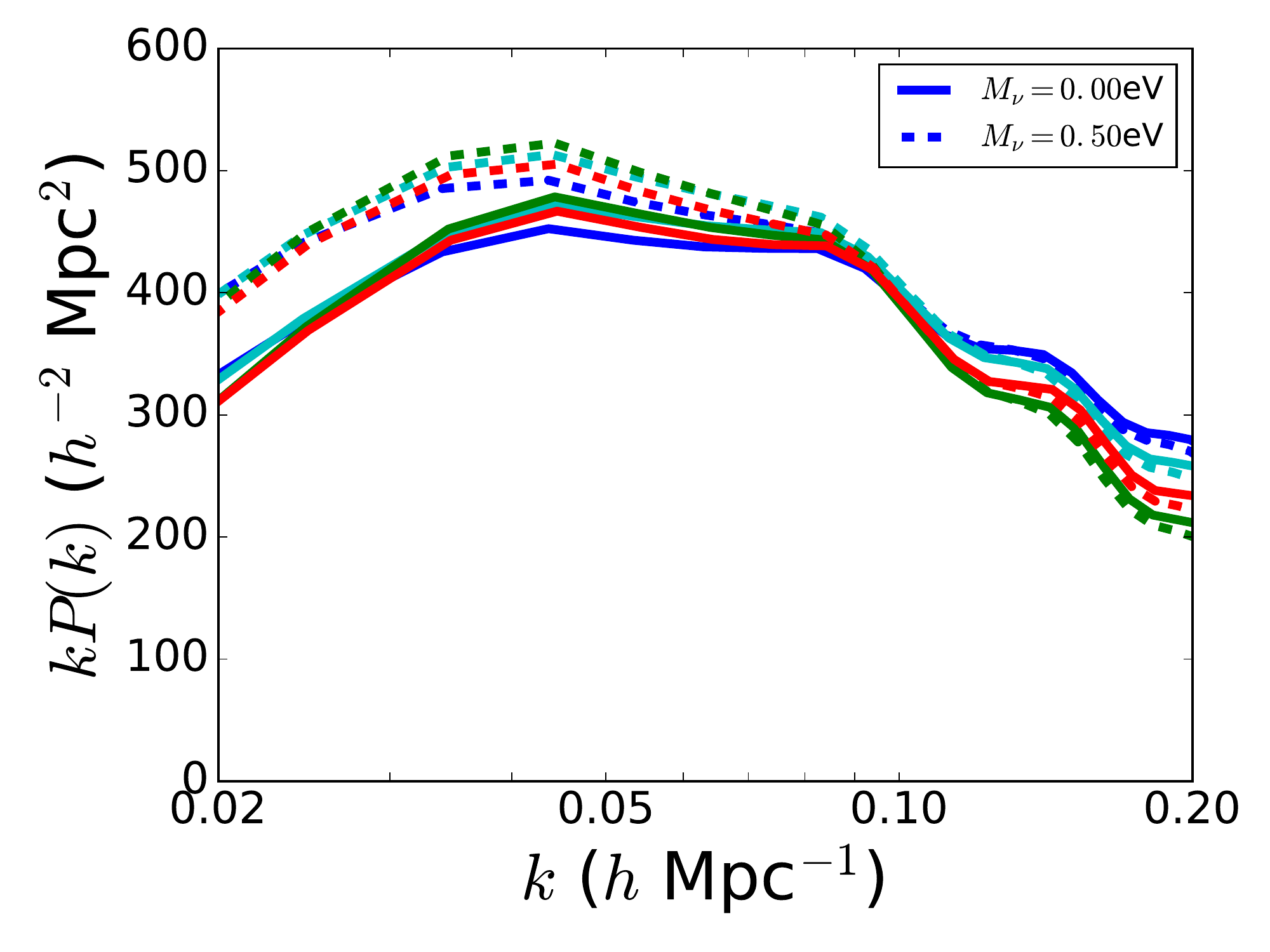}
\caption{LRG (left) and WZ (right) data sets used and the theoretical best-fit obtained from CMB+$P(k)$ data (solid line). The dashed line  corresponds to a model with $M_{\nu}$ corresponding to the 95\% limit for CMB data, keeping all other parameters fixed  except for $\omega_{\rm{cdm}}$. The WZ plot is normalised so that all the lines overlap at large scales. In the upper (lower) panels we show the non-linear matter power spectrum 
after (before) applying the window function of the two surveys. As opposed to Fig.~\ref{fig:pk_mnu}, here the theory power spectrum is multiplied by the marginalised value of the square of the galaxy bias for each model.} 
\label{fig:wz_lrg_pk_limits}
\end{figure}

In Table~\ref{tab:limits_15} and Table~\ref{tab:limits_15_lensing}, we report the limits at 95\%~C.L. on the sum of  neutrino 
masses (assuming, as standard, three active neutrinos) as obtained using recent  data from the Planck experiment  and Large Scale Structure measurements from WZ and LRG. 
 In Fig.~\ref{fig:wz_lrg_pk_limits} we show the effect of a non-zero neutrino mass on the non-linear matter power spectrum and the 
effect of applying the window function of the two surveys (compare the upper and lower panels). 

In Ref.~\cite{Ade:2015xua},  an upper bound of 
$M_\nu < 0.49~$eV (0.59~eV) was reported for the combination CMB15(+lensing). We find that adding the 
information from the LRG sample improves the limits to 0.38~eV (0.44~eV), while when considering the WZ data set, we find an 
upper limit of 0.37~eV (0.43~eV). 
The fact that the lensing information allows slightly higher values of the neutrino mass --
compare Table~\ref{tab:limits_15} with Table~\ref{tab:limits_15_lensing}-- was noticed already in the Planck analysis~\cite{Ade:2015xua} 
to which we refer for details and discussion. 

For the combination of CMB15+BAO, we find that, when adding the 
LRG information, the bound decreases to 0.13~eV, and to 0.17~eV if the CMB lensing information is also included. 
Considering, instead, the WZ information, we find compatible upper limits of 
0.14~eV, and 0.17~eV when the lensing information is also included. 
The constraints on the values of the remaining cosmological parameters allowed to vary  in our analysis 
 are summarised in  Table~\ref{tab:limits_lrg_wz_bao}.

\begin{figure}[h!]
\centering
\includegraphics[width=0.4\textwidth]{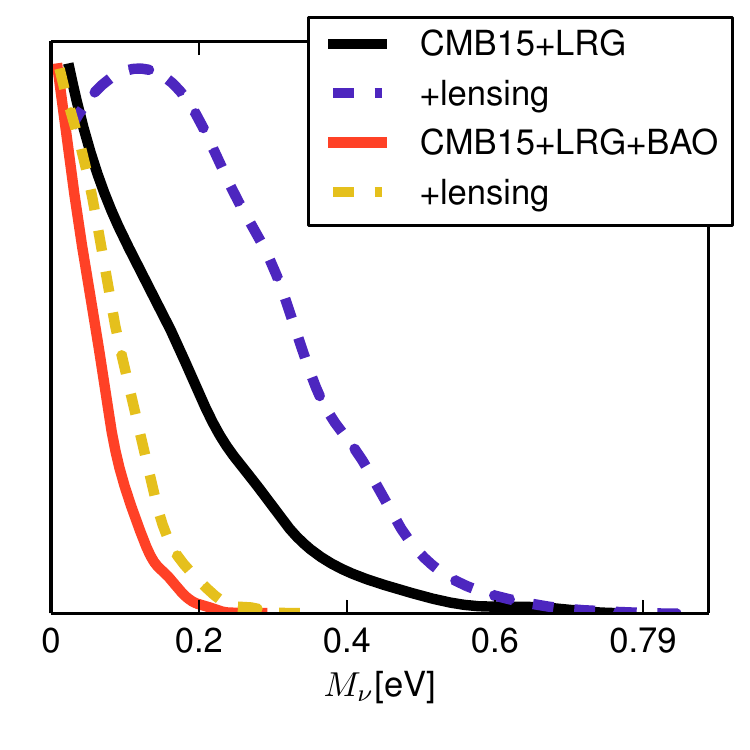} \hspace{0.5cm}
\includegraphics[width=0.4\textwidth]{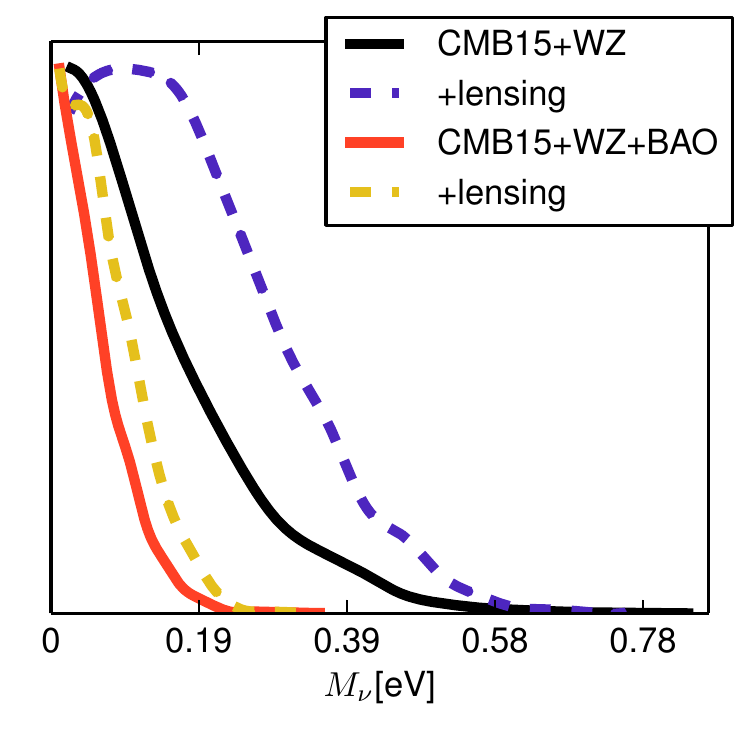} 
\caption{Comparison between the posterior distribution of the sum of neutrino masses, $M_\nu$, from the following data combinations: 
CMB15 + LRG (+lensing) and CMB15 + LRG + BAO (+lensing), in the left panel; CMB15 + WiggleZ (+lensing) and CMB15 + WiggleZ + BAO (+lensing), in the 
right panel. CMB15 indicates the 2015 Planck TT,TE,EE + lowP data. 
For a detailed description of the individual data sets used, we refer to Sec.~\ref{sec:data}. 
We use black (blue dashed) lines when power spectrum data are combined with CMB15 (+lensing) data and 
red (yellow dashed) lines when also BAO is added. 
}
\label{fig:1d-sdss-vs-wiggleZ}
\end{figure}

\begin{figure}[h!]
\centering
\includegraphics[width=0.7\textwidth]{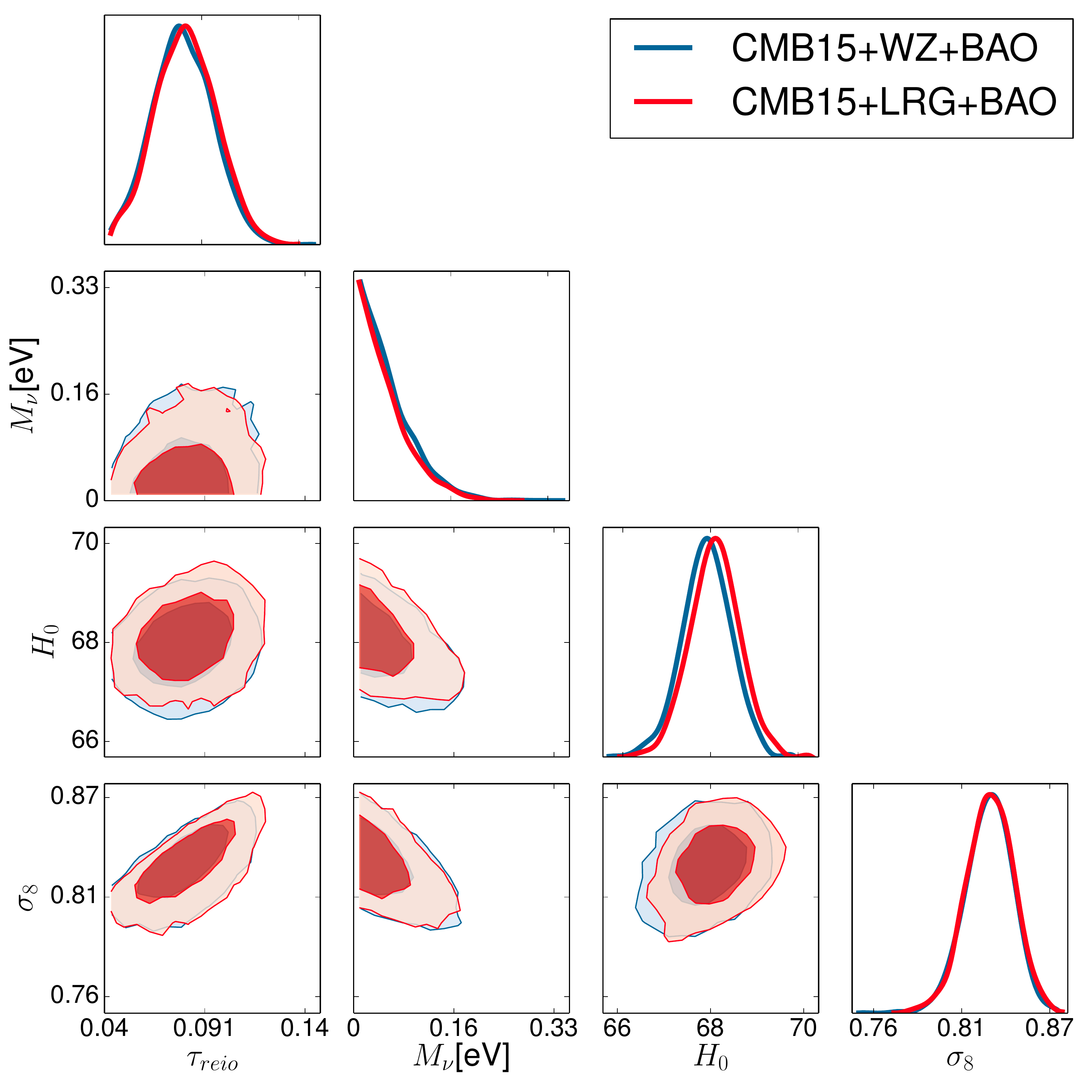}
\caption{Two-dimensional posterior distribution for ($\tau_{\rm{reio}}, M_\nu, H_0, \sigma_8$) parameters from 
the CMB15 + LRG + BAO (red contours) and CMB15 + WZ + BAO (blue contours) data sets. 
The $1~\sigma$ and $2~\sigma$ contours are shown.
}
\label{fig:2d-sdss-vs-wiggleZ}
\end{figure}

\begin{figure}[h!]
\centering
\includegraphics[width=0.4\textwidth]{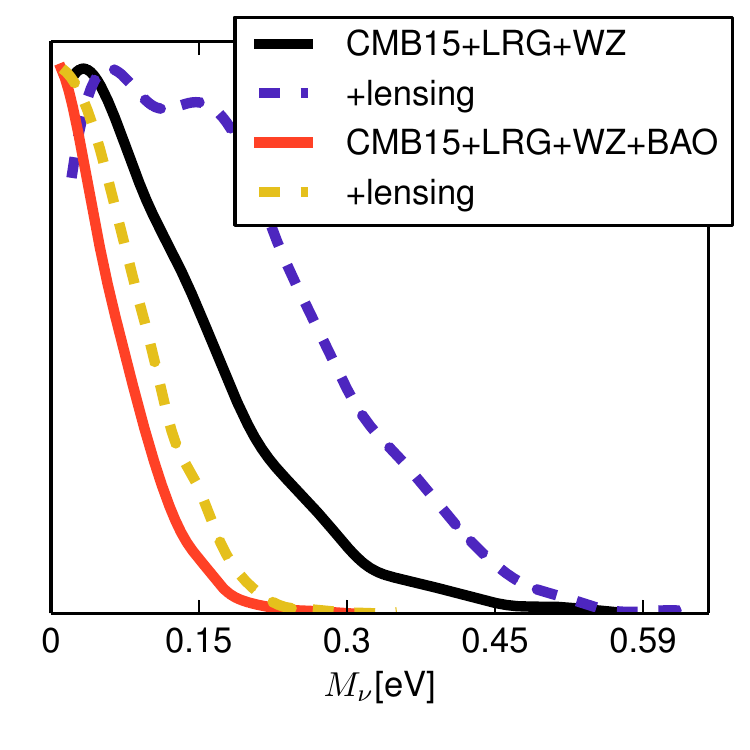}
\caption{Comparison between the posterior distribution of the sum of neutrino masses, $M_\nu$, from the following data combinations: 
CMB15 + LRG + WiggleZ (+lensing) and CMB15 + LRG + WiggleZ + BAO (+lensing). 
CMB15 indicates the 2015 Planck TT,TE,EE + lowP data. 
For a detailed description of the individual data sets used, we refer to Sec.~\ref{sec:data}. 
We use black (blue dashed) lines when power spectrum data are combined with CMB15 (+lensing) data and 
red (yellow dashed) lines when also BAO is added. 
}
\label{fig:1d-sdss-and-wiggleZ}
\end{figure}

\begin{figure}[h!]
\centering
\includegraphics[width=0.7\textwidth]{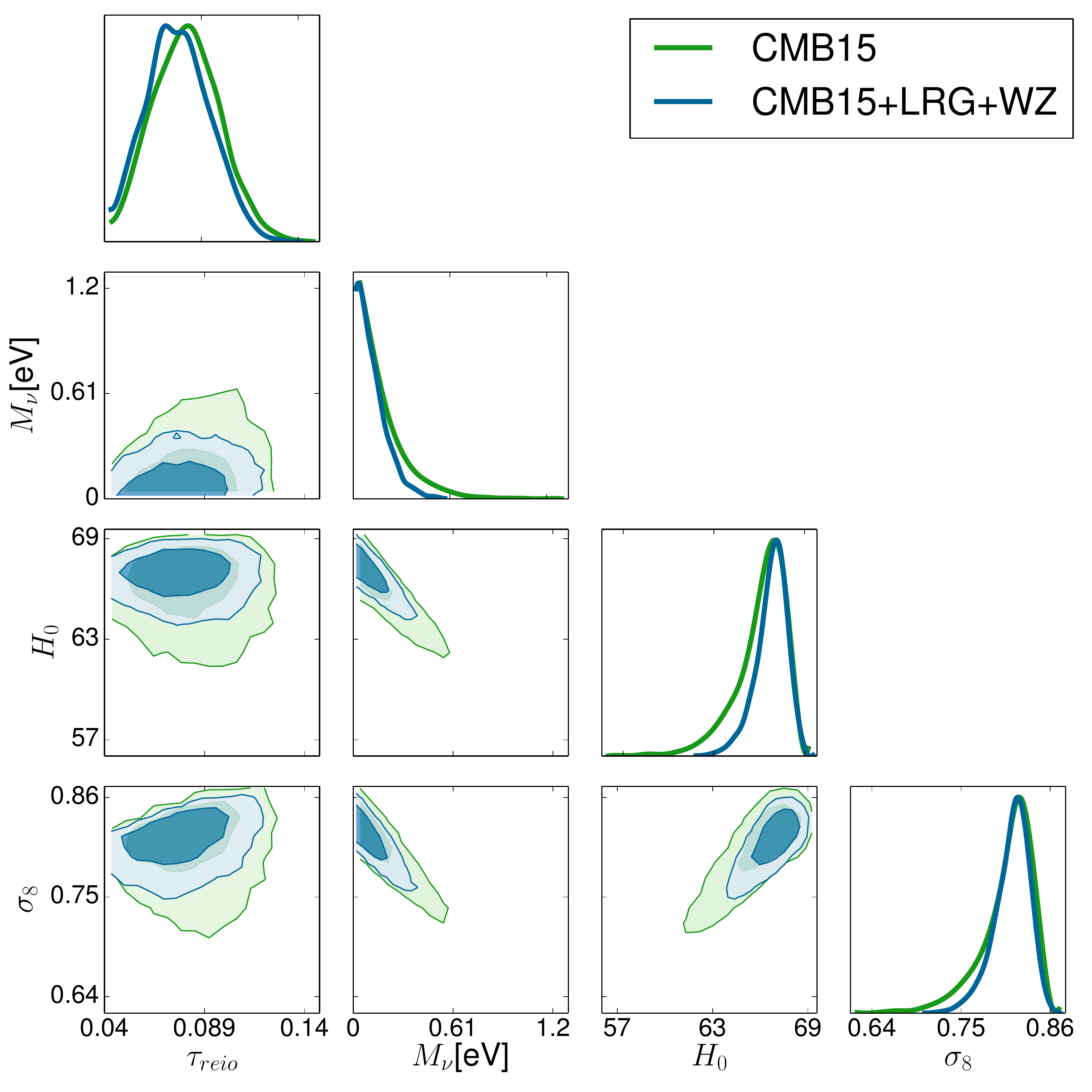}
\caption{Two-dimensional posterior distribution for ($\tau_{\rm{reio}}, M_\nu, H_0, \sigma_8$) parameters from 
the CMB15 (green contours) and CMB15 + LRG + WZ data sets (blue contours). The $1~\sigma$ and $2~\sigma$ contours are shown.
}
\label{fig:2d-cmb-sdss-and-wiggleZ}
\end{figure}

\begin{figure}[h!]
\centering
\includegraphics[width=0.7\textwidth]{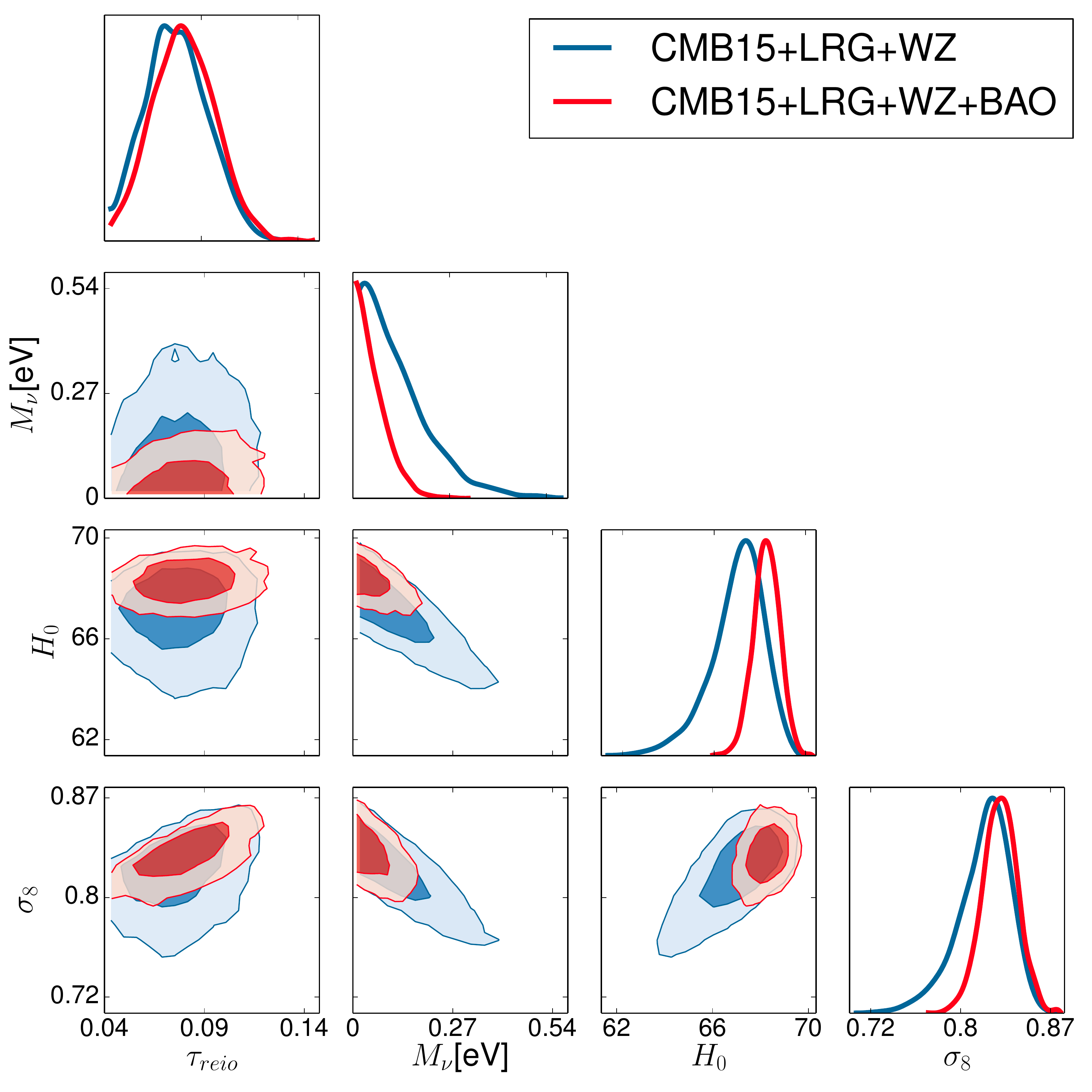}
\caption{Two-dimensional posterior distribution for ($\tau_{\rm{reio}}, M_\nu, H_0, \sigma_8$) parameters from 
the CMB15 + LRG + WZ + BAO (red contours) and CMB15 + LRG + WZ data sets (blue contours). The $1~\sigma$ and $2~\sigma$ contours are shown.
}
\label{fig:2d-sdss-and-wiggleZ}
\end{figure}

In Fig.~\ref{fig:1d-sdss-vs-wiggleZ}, we illustrate  the $M_{\nu}$ posterior distribution considering 
CMB15 combined with either LRG (left panel) or WZ (right panel). It can be clearly  seen  how  BAO information tightens the neutrino mass upper limits, providing a 
great improvement for both the LRG and the WZ data sets. 
  In order to clarify how future cosmological measurements can help break further the degeneracies in neutrino mass measurements,  we report the likelihood contours for all
the relevant cosmological parameters  ($M_{\nu}$, $\tau_{\rm{reio}}$, $H_0$, and $\sigma_8$) for the combination of CMB15+LRG+BAO and CMB15+WZ+BAO in Fig.~\ref{fig:2d-sdss-vs-wiggleZ}. 
 The effect of combining both galaxy surveys is shown in Fig.~\ref{fig:1d-sdss-and-wiggleZ}, Fig.~\ref{fig:2d-cmb-sdss-and-wiggleZ} and 
 Fig.~\ref{fig:2d-sdss-and-wiggleZ},  and  the $M_{\nu}$ bounds are reported in Table~\ref{tab:limits_15} and in 
Table~\ref{tab:limits_15_lensing}. 

\begin{table}
\centering
\begin{tabular}{ c || c } 
Data sets & \multicolumn{1}{c}{$M_\nu$ at 95\% CL}  \\ \hline \hline
CMB15 + LRG & 0.38 eV \\\hline
CMB15 + WZ & 0.37 eV  \\\hline
CMB15 + LRG + WZ & 0.30 eV \\\hline \hline
 CMB15 + LRG + BAO &  0.13 eV\\ \hline 
 CMB15 + WZ + BAO &  0.14 eV \\ \hline 
CMB15 + LRG + WZ + BAO & 0.14 eV \\\hline
\end{tabular}
\caption{Upper bounds (at 95\%~C.L.) on the sum of the three active neutrino masses, 
$M_\nu$, 
for different combinations of data sets. With CMB15 we indicate the 
combination of Planck TT,TE,EE + lowP data, while  
LRG and WZ designate the full shape of the matter power spectra as reported by 
SDSS-DR7 and WiggleZ survey respectively.
For the detailed description of each single data set used, we refer to Sec.~\ref{sec:data}. 
}
\label{tab:limits_15}
\end{table}

\begin{table}
\centering
\begin{tabular}{ c || c } 
Data sets & \multicolumn{1}{c}{$M_\nu$ at 95\% CL}  \\ \hline \hline
CMB15 + LRG + lensing & 0.44 eV\\\hline
CMB15 + WZ + lensing & 0.43 eV\\\hline
CMB15 + LRG + WZ + lensing & 0.38 eV \\\hline \hline
CMB15 + LRG + BAO + lensing & 0.17 eV\\ \hline 
CMB15 + WZ + BAO + lensing & 0.17 eV\\ \hline 
CMB15 + LRG + WZ + BAO + lensing & 0.18 eV\\\hline
\end{tabular}
\caption{Same as Table~\ref{tab:limits_15}, using also the CMB lensing measurement. 
}
\label{tab:limits_15_lensing}
\end{table}

\begin{table}
\centering
\begin{tabular}{ c || c | c } 
Parameter & CMB15+LRG+BAO & CMB15+WZ+BAO  \\ \hline \hline 
$100~\omega_b$ & 2.236$_{-0.014}^{+0.014}$ & 2.233$_{-0.015}^{+0.015}$ \\\hline 
$\omega_{\rm{cdm}}$ & 0.1183$_{-0.0011}^{+0.0012}$ & 0.1186$_{-0.0011}^{+0.0011}$ \\\hline 
$n_s$ & 0.9677$_{-0.0045}^{+0.0042}$ & 0.9669$_{-0.0044}^{+0.0040}$ \\ \hline
$\tau_{\rm{reio}}$ & 0.083$_{-0.017}^{+0.016}$ & 0.081$_{-0.017}^{+0.016}$ \\ \hline
$ln(10^{10}A_{s})$  &$3.097_{-0.034}^{+0.031}$ & $3.095_{-0.033}^{+0.031}$ \\ \hline
$H_0$ & 68.06$_{-0.55}^{+0.55}$ & 67.89$_{-0.52}^{+0.56}$\\ \hline
$\sigma_8$ & 0.831$_{-0.015}^{+0.016}$ & 0.830$_{-0.014}^{+0.016}$ \\ \hline  
$M_\nu$~[eV] & $<0.13$ & $<0.14$ \\\hline
\end{tabular}
\caption{Constraints on the values of the cosmological parameters from 
CMB15+LRG+BAO and CMB15+WZ+BAO. 
With CMB15 we indicate the 
combination of Planck TT,TE,EE + lowP data, while  
LRG and WZ designate the full shape of the matter power spectra as reported by 
SDSS-DR7 and WiggleZ survey respectively.
For the detailed description of the data set used, we refer to Sec.~\ref{sec:data}. 
For the total neutrino mass $M_\nu$ we report the 95\%~C.L. upper limit, whereas for all the other parameters  we quote the mean values 
and $1~\sigma$  error bars. 
}
\label{tab:limits_lrg_wz_bao}
\end{table}
 
We can conclude that, if the BAO information is not included,  LRG and WZ provide a similar 
upper limit on the sum of  neutrino masses. Considering the information from the two galaxy surveys together improves the constraints 
on $M_\nu$. 

Our results for the CMB15+LRG+BAO data set combination are competitive with the strongest upper limit on neutrino masses presented in the 
literature, that is the value of $M_{\nu}< 0.12$~eV at $95\%$ C.L.~\cite{Palanque-Delabrouille:2015pga}, obtained 
using Lyman-$\alpha$ data from  the BOSS survey  together with the recent Planck~2015  release and BAO data. 
This is an important result that shows how cosmology is pointing towards a
similar value for the upper limit on the sum of neutrino masses $M_\nu$, independently of the 
large-scale structure  probe considered. Poorly known astrophysical processes  are the main limitation in extracting constraints on neutrino masses from large-scale structures.
Different tracers of the large scale structure are affected by different  astrophysical processes: the fact that  approaches as different as the one of  \cite{Palanque-Delabrouille:2015pga} and the one presented here find consistent results, supports the robustness of the reported limit.

We find a very weak (if any at all) correlation between the parameter $\tau_{\rm{reio}}$ and $M_{\nu}$.
On the other hand Fisher-based forecasts for future data, see e.g., \cite{Liuetal15, Allisonetal15} find a correlation between the two parameters albeit of different degree and orientation depending on the  specific experimental set up. It seems that stronger constraints on the expansion history than those adopted here   are responsible for creating a correlation between these two parameters. In fact we can approximately  reproduce the  correlation found by \cite{Allisonetal15} if we consider  an extended data set including CMB15+BAO, the latest supernovae data compilation  JLA \cite{JLA}
 and the $H_0$ determination \cite{Riess2011,Humphreys2013,Cuesta15}. The degeneracy arises  because of  the $A_s$ vs $\exp(-2\tau)$ CMB degeneracy, which is partly  broken by polarisation information;  the remaining  degeneracy  however propagates through $M_{\nu}$. 
 Here we have decided not to combine together  too many data sets, and thus  we do not find any significant  correlation. It remains to be  explored whether for future data, adding information on the power spectrum shape coming from large-scale structure data or by further constraining $\tau_{\rm{reio}}$ e.g., \cite{Liuetal15,Allisonetal15,DiValentino:2015sam}, would break the $\tau_{\rm{reio}}-M_{\nu}$ degeneracy. 
In Appendix A we report the neutrino mass constraints if we also add the  local determination of $H_0$~\cite{Riess2011,Humphreys2013,Cuesta15}. Since a tension has been claimed between the CMB and $H_0$ measurements within the standard $\Lambda$CDM model and its simple extensions, we will not dwell further on these limits here, but report them for completeness. 

\begin{figure}[t!]
\centering
\includegraphics[width=0.8\textwidth , height=12cm]{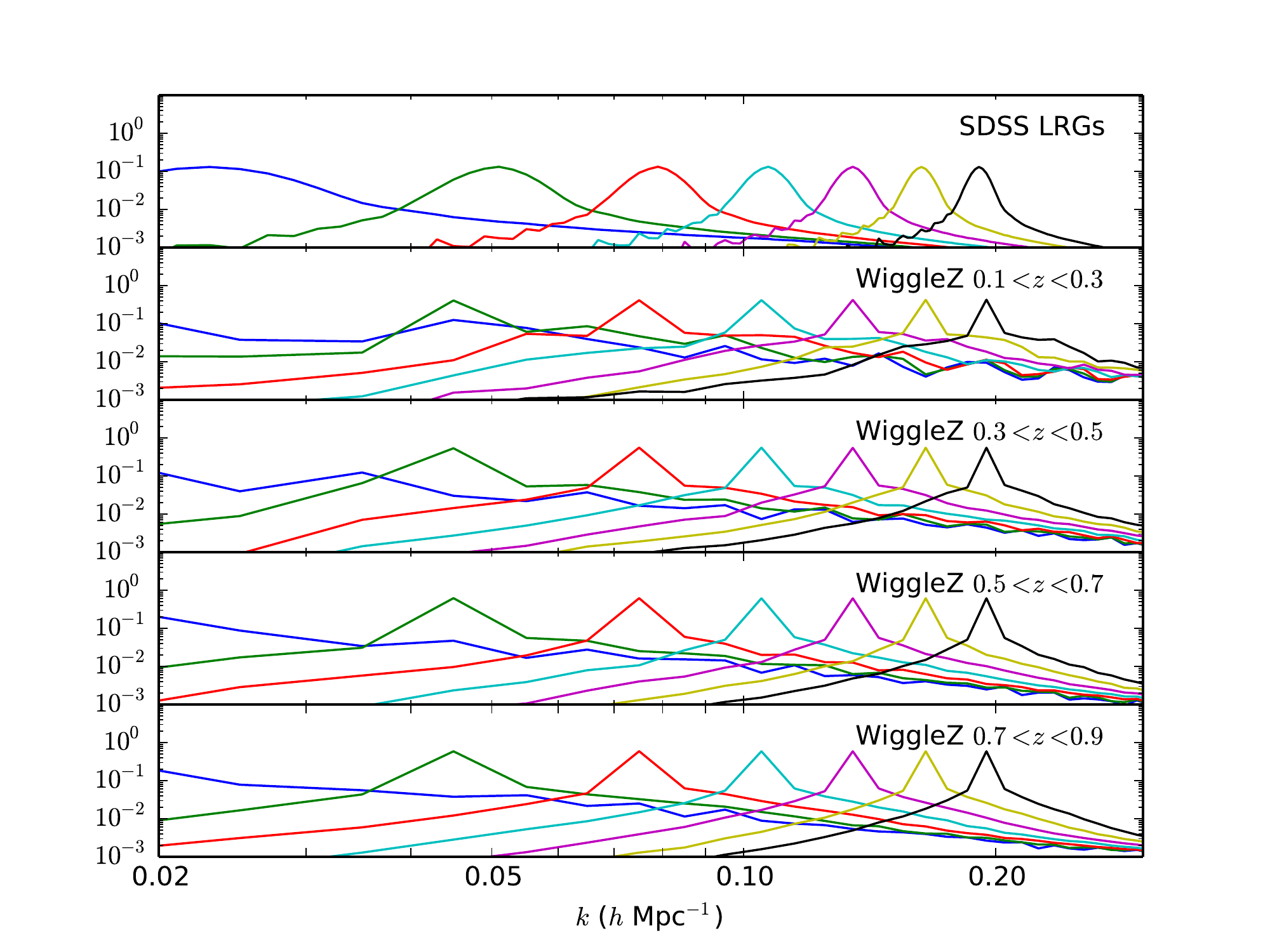}
\caption{Window functions for a sample of $k$-bins, for LRG  (top panel) and the four redshift bins of WZ (bottom four panels).}
\label{fig:windows}
\end{figure}

It is interesting to  compare the geometries of the two  galaxy surveys considered and the possible effect on the constraints. 
WZ covers a larger volume  (the effective volume   of WZ is  $V_{\rm eff} =0.34  h^{-3}$Gpc$^3$ and LRG cover  $V_{\rm eff} =0.26  h^{-3}$Gpc$^3$ ).  Since the errors on the power spectrum measurements scale like $V_{\rm eff}^{-1/2}$  for this argument WZ is expected to produce slightly tighter constraints in the ideal case where the best fit parameter values coincide.  Because one must marginalise over an (unknown) galaxy bias amplitude, the sensitivity of large-scale structure data to   $M_{\nu}$  is due to a measurement of the broadband shape of the power spectrum  over the scales probed by the galaxy surveys. Here is where  a narrow  survey window function helps. The survey window  function  couples Fourier modes, transferring  power from $k$ to $k'$ thus smoothing out features. For $M_{\nu}$ of 0.1 eV the power suppression (compared to the massless case) is $\sim 3.5$\% at $k=0.02 h$/Mpc and $\sim 6$\% at $k=0.2h$/Mpc i.e., a $\sim 2$\% effect.
For reference, over the same  $k$-range, the matter $P(k)$ drops by more than an order of magnitude, so  even a  transfer of $\sim 0.1$\% of power between these two scales completely swamps the signal.  In Fig.~\ref{fig:windows} we show the window functions of LRG and WZ where it is possible to appreciate that the WZ window functions have broader wings than the LRG ones. This can be understood  by considering the two surveys geometry: LRG covers  $7931$  square degrees of which  $7190$  are contiguous --and three redshift bins combined into a single redshift measurement of $P(k)$-- while WZ covers 1000 square degrees divided in 7 disconnected patches and 4 redshift bins.

We have decided  not to include in our analysis results  from galaxy weak lensing surveys, like CFHTLens~\cite{Heymans:2012gg,Erben:2012zw}, 
since a tension is present with the base $\Lambda$CDM model (see discussion in~\cite{Leistedt2014,Ade:2015zua}). This discrepancy might be partially alleviated by extended models, with massive  neutrinos and varying $N_{\rm eff}$ (for other analysis in this direction, using also data from Redshift Space Distortion 
and Baryon Acoustic Oscillation information, we refer to Ref.~\cite{Beutler:2014yhv} and Ref.~\cite{Samushia:2013yga}) 
However, it was concluded in Ref.~\cite{Leistedt2014, Ade:2015zua} that modifications 
to the neutrino sector alone cannot resolve the  tension and that further investigation into  systematic effects or interpretation of the data is needed. For this reason, we decided to use the  full shape information on the matter power spectrum as probed by galaxy redshift surveys instead. 
 
\section{\label{sec:conc}Conclusions}
It has long be recognised that cosmology  can provide key information to learn about neutrino physics, most notably by  constraining the absolute neutrino mass scale.  However it is only very recently that the accuracy of cosmological measurements  has reached a level where it could be used to determine the hierarchy and/or  detect the effect of neutrino masses on the evolution of the universe  and  thus  provide a measurement of the sum of the masses $M_{\nu}$.  In fact oscillation experiments impose a minimum $M_{\nu}$ for the inverted hierarchy around $M_{\nu}\gtrsim 0.1$ eV (e.g.,\cite{JKPGV}),  and cosmological bounds are  now reaching this level of accuracy. However accuracy is not enough:  for cosmological constraints, to reach this level of information about the clustering of large-scale  structure, tracers such as the Lyman-alpha forest or galaxies must be included. These tracers are believed to be affected by poorly known astrophysics which makes them biased tracers with a  poorly known bias. Thus these constraints should be shown to be also robust.

In this paper, we have analysed the robustness of the neutrino mass upper limit under different 
combinations of data sets. In particular, we have focused our analysis on the 
constraints from different  galaxy power spectrum data. We have analysed in detail the information provided by two data sets, 
namely the WiggleZ Dark Energy Survey (WZ) and the Luminous Red Galaxies (LRG) sample of 
the SDSS-DR7.  We have combined the information on the full shape of the matter power spectrum from 
these two data sets with the recent data from the Planck~2015 release and state-of-the-art BAO data. 

The analysis presented here   serves two purposes. Firstly we have demonstrated the level of robustness of the neutrino mass constraints obtained including galaxy clustering information.  
 LRG galaxies are  thought to inhabit preferentially  the centre of cluster-size dark matter halos  which populate high-density regions while WZ ones tend instead  to avoid the densest regions. Thus not only the bias of the two tracers are very different but also the sensitivity of their power spectra to non-linearities.
 When including BAO data which reduce cosmological degeneracies present in CMB constrains,  the LRG data set yields 
 an upper limit of 0.13~eV at 95\% C.L., while  WZ results in an upper limit of 0.14~eV. 
Thus the LRG and WZ provide similar constraints, offering confidence that  systematic effects due to galaxy bias or non-linearities are  below the error bars for the respective limits.
 
Secondly we note that the LRG limit  ($M_{\nu}<$0.13~eV) is  very close  to the tightest limit in the 
literature, that corresponds to 0.12~eV  obtained  using the Lyman-$\alpha$ 
power spectrum from BOSS~\cite{Palanque-Delabrouille:2015pga}. The  Lyman-$\alpha$  signal  corresponds to a yet completely different clustering  tracer, mapping out the neutral Hydrogen in  intergalactic medium. This further supports the robustness  of the reported limit. 
This limit is reaching tantalisingly close to the 0.1 eV target corresponding to roughly the minimum $M_{\nu}$ allowed  for  active neutrinos with inverted hierarchy. A detailed discussion of the interpretation of the  Lyman-$\alpha$ $M_{\nu}$ constraint  also in light of  neutrino experiments is presented in~\cite{Dell'Oro:2015tia}. Needless to say that very similar   considerations apply to  the result  presented here.

 It was described in  e.g., Ref.~\cite{Strumia:2005tc, Haba:2013xwa} how the constraints on $M_\nu$ can provide information on the neutrino 
mass degeneracy. The authors concluded that a bound of $M_\nu < 0.23$~eV was able to reject at 95\% C.L. neutrino mass degeneracy larger than 
85\% (82.5\%) for normal (inverted) hierarchy. The bounds of 0.13~eV and 0.14~eV found in this work further support 
the conclusion of Ref.~\cite{Haba:2013xwa} and the constraints on the magnitude of degeneracy for neutrino masses.  Moreover,  Fig.~3 of Ref.   \cite{Haba:2013xwa}   interpreted in light of the present constraint, indicates that neutrinoless double beta decay experiments  should strive to reach   ton-size detectors.  
 
 As a consequence, only a very small region of parameter space for the inverted hierarchy is still allowed, and even a modest improvement in current constraints could determine the hierarchy. Conversely, if the mass hierarchy is the inverted one, then a detection of neutrino mass from the sky and determination of the mass scale is just around the corner.

In conclusion, the information on the full shape of the matter power spectrum,  accessible from large-volume large-scale structure  surveys (possibly with with narrow window functions), is becoming of fundamental importance for  neutrino physics  and 
in particular in shedding light on the mass scale and the hierarchy.

\section*{\label{sec:ack}Acknowledgements}
We acknowledge Signe Riemer-S{\o}rensen for discussions on the WiggleZ data analysis. 
LV and AJC are supported by the European Research Council under the European Community's Seventh Framework Programme FP7-IDEAS-Phys.LSS 240117. 
Funding for this work was partially provided by the Spanish MINECO under projects AYA2014-58747P and MDM-2014-0369 of ICCUB (Unidad de Excelencia 'Mar{\'\i}a de Maeztu').
AJC  acknowledges hospitality of ITC, Harvard--Smithsonian Center for Astrophysics, Harvard University.
VN acknowledges support by Spanish MINECO through project FPA2012-31880, by Spanish MINECO (Centro de excelencia Severo Ochoa Program) under grant SEV-2012-0249 and 
by the European Union through the FP7 Marie Curie Actions ITN INVISIBLES (PITN-GA-2011-289442). The work of VN was also supported by the Deutsche
Forschungsgemeinschaft (DFG) through the Collaborative Research Centre SFB 676 ``Particles, Strings and the Early Universe'' (through an SFB fellowship). 
VN would like to express a special thanks to the Mainz Institute for Theoretical Physics (MITP) for its 
hospitality and support. 

Based on observations obtained with Planck (\url{http://www.esa.int/Planck}), an ESA science mission with instruments and contributions directly funded by ESA Member States, NASA, and Canada.

Funding for the SDSS and SDSS-II has been provided by the Alfred P. Sloan Foundation, the Participating Institutions, the National Science Foundation, the U.S. Department of Energy, the National Aeronautics and Space Administration, the Japanese Monbukagakusho, the Max Planck Society, and the Higher Education Funding Council for England. The SDSS Web Site is \url{http://www.sdss.org/}. 

The SDSS is managed by the Astrophysical Research Consortium for the Participating Institutions. The Participating Institutions are the American Museum of Natural History, Astrophysical Institute Potsdam, University of Basel, University of Cambridge, Case Western Reserve University, University of Chicago, Drexel University, Fermilab, the Institute for Advanced Study, the Japan Participation Group, Johns Hopkins University, the Joint Institute for Nuclear Astrophysics, the Kavli Institute for Particle Astrophysics and Cosmology, the Korean Scientist Group, the Chinese Academy of Sciences (LAMOST), Los Alamos National Laboratory, the Max-Planck-Institute for Astronomy (MPIA), the Max-Planck-Institute for Astrophysics (MPA), New Mexico State University, Ohio State University, University of Pittsburgh, University of Portsmouth, Princeton University, the United States Naval Observatory, and the University of Washington.

Funding for SDSS-III has been provided by the Alfred P. Sloan Foundation, the Participating Institutions, the National Science Foundation, and the U.S. Department of Energy Office of Science. The SDSS-III web site is \url{http://www.sdss3.org/}.

SDSS-III is managed by the Astrophysical Research Consortium for the Participating Institutions of the SDSS-III Collaboration including the University of Arizona, the Brazilian Participation Group, Brookhaven National Laboratory, Carnegie Mellon University, University of Florida, the French Participation Group, the German Participation Group, Harvard University, the Instituto de Astrofisica de Canarias, the Michigan State/Notre Dame/JINA Participation Group, Johns Hopkins University, Lawrence Berkeley National Laboratory, Max Planck Institute for Astrophysics, Max Planck Institute for Extraterrestrial Physics, New Mexico State University, New York University, Ohio State University, Pennsylvania State University, University of Portsmouth, Princeton University, the Spanish Participation Group, University of Tokyo, University of Utah, Vanderbilt University, University of Virginia, University of Washington, and Yale University.

\clearpage

\appendix
\section{Including $H_0$ determination}
We report (Table~\ref{tab:limits_all})  constraints  on $M_{\nu}$ when including also the $H_0$ determination \cite{Riess2011,Humphreys2013,Cuesta15} of $H_0=73.0\pm 2.4$ km s$^{-1}$ Mpc$^{-1}$.
\begin{table}[!h]
\centering
\begin{tabular}{ c || c } 
Data sets & \multicolumn{1}{c}{$M_\nu$ at 95\% CL}  \\ \hline \hline
CMB15 + BAO + $H_0$ & 0.11 eV\\ \hline
CMB15 + LRG + $H_0$ &  0.19 eV\\
CMB15 + WZ + $H_0$ & 0.18 eV \\
CMB15 + LRG + WZ + $H_0$ & 0.175 eV \\
CMB15 + LRG + BAO + $H_0$ & 0.11 eV\\
CMB15 + WZ + BAO + $H_0$ &0.11 eV \\
CMB15 + LRG + WZ + BAO + $H_0$ & 0.12 eV \\ \hline
CMB15 + LRG + lensing + $H_0$ & 0.23 eV\\
CMB15 + WZ + lensing + $H_0$ &  0.21 eV\\
CMB15 + LRG + WZ + lensing + $H_0$ & 0.21 eV\\
CMB15 + LRG + BAO + lensing + $H_0$ & 0.135 eV\\ 
CMB15 + WZ + BAO + lensing + $H_0$& 0.14  eV\\ 
CMB15 + LRG + WZ + BAO + lensing + $H_0$& 0.14 eV\\ \hline
\end{tabular}
\caption{Same as Table~\ref{tab:limits_15} and  Table~\ref{tab:limits_15_lensing} using also the $H_0$ determination of \cite{Riess2011,Humphreys2013,Cuesta15}. 
}
\label{tab:limits_all}
\end{table}

\bibliographystyle{./apsrev}
\bibliography{nus}

\end{document}